\newcommand{\NDMAP}{Ni(C$_5$D$_{14}$N$_2$)$_2$N$_3$(PF$_6$)}
\newcommand{\NENP}{Ni(C$_2$H$_8$N$_2$)$_2$NO$_2$(ClO$_4$)}
\newcommand{\NDMAZ}{Ni(C$_5$H$_{14}$N$_2$)$_2$N$_3$(ClO$_4$)}
\newcommand{\NTENP}{[Ni($N,N'$-bis(3-aminopropyl)propane-1,3-diamine($\mu$-NO$_2$)]ClO$_4$}

\documentclass[
    ,final            
  ]
  {aipproc}

\layoutstyle{8x11double}

\begin{document}

\title{Neutron Scattering Study of Quantum Phase Transitions in
Integral Spin Chains.}



\author{A. Zheludev}{
  address={Condensed Matter Sciences Division, Oak Ridge
National Laboratory, Oak Ridge, TN 37831-6393, USA.} }

\begin{abstract}
Quite a few low-dimensional magnets are quantum-disordered ``spin
liquids'' with a characteristic gap in the magnetic excitation
spectrum. Among these are antiferromagnetic chains of integer
quantum spins. Their generic feature are long-lived massive
(gapped) excitations (magnons) that are subject to Zeeman
splitting in external magnetic fields. The gap in one of the
magnon branches decreases with field, driving a soft-mode quantum
phase transition. The system then enters a qualitatively new
high-field phase. The actual properties at high fields,
particularly the spin dynamics, critically depend on the system
under consideration. Recent neutron scattering studies of
organometallic polymer crystals NDMAP (Haldane spin chains with
anisotropy) and NTENP (dimerized $S=1$ chains) revealed rich and
unique physics.

\end{abstract}

\maketitle


\section{Introduction}
A quarter of a century ago the interest in quantum magnetism was
rekindled by the famous work of Haldane \cite{Haldane83}, who
showed that integer-spin one-dimensional (1D) Heisenberg
antiferromagnets (AFs) have a non-degenerate ground state and an
energy gap in the excitation spectrum. Quantum zero-point
fluctuations are so strong that the spin correlation function
decays exponentially, and these systems  can be viewed as 1D
``spin liquids''. Their properties are in stark contrast with
those of half-integer spin chains that are gapless, and where spin
correlations decay according to a slow power-law (Luttinger spin
liquds).

\begin{figure}
 \includegraphics[width=3in]{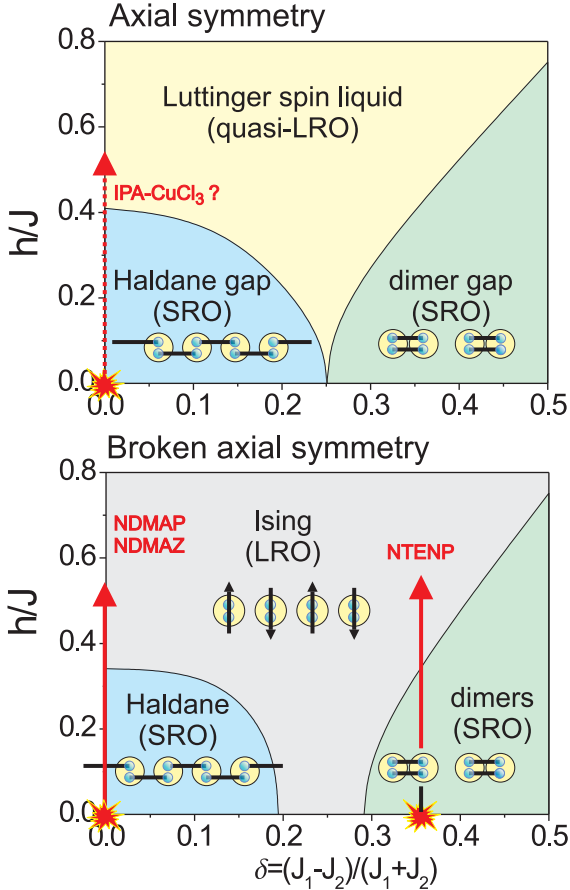}
 \caption{\label{phase} Schematic phase diagrams of bond-alternating $S=1$ chains in external
 magnetic fields for axially symmetric (top) and axially asymmetric (bottom) models.}
\end{figure}

Haldane spin chains have been studied extensively, and are by now
very well understood. The exact ground state is not known, but is
similar to the easy to visualize Valence Bond Solid (VBS) state
\cite{Affleck89}. The latter is constructed by representing each
$S=1$ spin as two separate $S=1/2$ spins, binding pairs of these
into antiferromagnetic dimers for each exchange bond, and
projecting the resulting state back onto the subspace where
$S_i^2=2$ on each site, as schematically shown in the insets on
the left side Fig.~\ref{phase}. Each exchange link carries exactly
one valence bond, and the periodicity of the underlying crystal
lattice remains intact. Considerably less attention has been given
to a {\it different} quantum spin liquid that is realized in $S=1$
AF spin chains with  exchange interactions of alternating
strength. As the alternation parameter
$\delta=(J_1-J_2)/(J_1+J_2)$ deviates from zero, the energy gap
$\Delta$ decreases and closes at some critical value
$|\delta|=\delta_c\approx 0.26$
\cite{Affleck87-2,Yamamoto94,Kitazawa96,Kohno98}. Increasing
$|\delta|$ beyond this quantum-critical point re-opens the spin
gap. The ground state is then no longer the Haldane state, but a
dimerized one. The corresponding valence bond wave function is
shown in right inset in Fig.~\ref{phase}. It contains two valence
bonds on each strong link and none on the weaker ones.

The two gapped quantum phases differ by their ``hidden'' spin
correlations \cite{Kennedy92,Yamamoto94}. The Haldane state has  a
non-vanishing AF string order parameter \cite{Nijs89} related to a
breaking of a non-local $Z_2\times Z_2$ symmetry \cite{Kennedy92}.
The highly non-local multi-spin correlation function that defines
AF strings  can not be expressed through the usual pair spin
correlation functions $\langle S_i(0) S_j(t)\rangle$, that
contribute to the dynamic susceptibility
$\chi(\mathbf{q},\omega)$. As a result, string order can not be
directly observed experimentally.

Qualitatively new physics is expected to emerge in strong magnetic
fields. Since the lowest-energy gap excitations are a $S=1$
triplet, they are subject to Zeeman splitting. With increasing
field gap in one of the three modes decreases and eventually
vanishes, leading to a 1-D Bose condensation of magnons
\cite{Schulz86,Katsumata89,Affleck90,Takahashi91}. A qualitatively
new state emerges at higher fields. For isotropic or XY-like spin
chains (axial symmetry) theory predicts an extended
quantum-critical state with power-law spin correlations and
certain incommensurate features
\cite{Tsvelik90,Affleck90,Affleck91,Takahashi91,Sachdev94,Furusaki99}.
However, in the presence of magnetic anisotropy the high-field
state is magnetically ordered at $T=0$: a {\it uniform} external
filed suppresses quantum spin fluctuations and induces {\it
staggered} (AF) long-range order \cite{Tsvelik90,Affleck90}. Even
in the ordered state quantum effects are significant, and the
system is a ``quantum spin solid''. Schematic phase diagrams for
the axially symmetric and asymmetric cases are shown in
Fig.~\ref{phase}.

Field-induced condensation of magnons and long-range ordering was
studied in a number of quantum-disordered gapped spin systems. The
best known examples are those of the spin-Peierls material
CuGeO$_3$ \cite{Kiryukhin95}, the $S=1/2$ Shastry-Sutherland
lattice SrCu$_2$(BO$_3$)$_2$ \cite{Kageyama99,Onizuka2000} and the
dimer compound TlCuCl$_3$ \cite{Ruegg2003}. None of these
materials realize the $S=1$ model, and all fall far from being
ideal 1D magnets. Moreover, for CuGeO$_3$ \cite{Kiryukhin95} and
SrCu$_2$(BO$_3$)$_2$ \cite{Kodama2002} the physics is complicated
by the involvement of the crystal lattice. The present review is
focused on a plain Heisenberg $S=1$ chain, where it is the
integrity of spin and the non-trivial topology of one dimension
that conspire to produce spectacular and exotic behavior in high
fields. The central question that will be asked is whether
Haldane-gap and dimerized $S=1$ chains behave any differently when
magnetized?

Excellent realizations of the $S=1$ Heisenberg chain model can be
found among Ni-based organometallic polymer crystals. In this
family \NENP~(NENP) is perhaps the one most extensively studied
materila \cite{Renard87,Ma92,Regnault93,Regnault94,Zaliznyak98}.
Unfortunately, the magnetic Ni$^{2+}$ ions in NENP have a
staggered $g$-factor \cite{Chiba91,Mitra94}. As a result, an
external magnetic field necessarily induces an internal staggered
field that ruins the phase transition. A breakthrogh in the study
of high-field quantum phases of $S=1$ chains came with the recent
discovery of the compounds \NDMAP~(NDMAP)
\cite{Honda98,Honda99,Chen2001,Zheludev2002,Zheludev2003,Hagiwara2003,Zheludev2004,Zheludev2005}
and \NDMAZ~(NDMAZ) \cite{Honda97,Zheludev2001-2}, that are
structurally similar to NENP, but are free of the staggered
$g$-factor problem. An even more exciting development was the
discovery of \NTENP~(NTENP)
\cite{Narumi2001,Zheludev2004-2,Regnault2004,Hagiwara2005}, that
is also structurally similar, but features bond-alternating $S=1$
chains. In the study of high-field phenomena in NDMAP, NDMAZ and
NTENP neutron scattering proved itself as the most useful
experimental tool.

\section{Dimerized vs. Haldane chains}
\begin{figure}
 \includegraphics[width=3in]{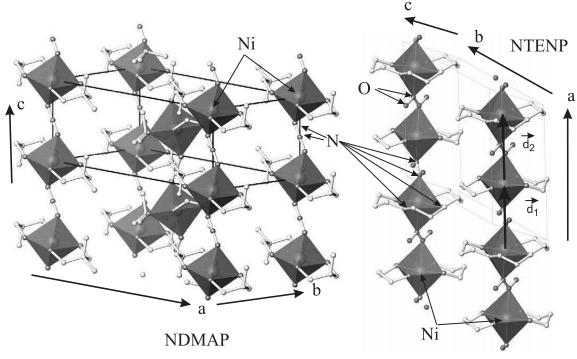}
 \caption{\label{struc} A schematic view of the uniform $S=1$ Ni$^{2+}$ spin chains
 in NDMAP and bond-alternating  chains  in NTENP. The equatorial vertices of the Ni$^{2+}$ coordination
 octahedra are nitrogen atoms. The octahedra are coupled via azide and NO$_2$ groups.
 Oxygen and nitrogen atoms are shown as dark grey spheres. Light grey spheres are carbon atoms.}
\end{figure}

\begin{figure}
 \includegraphics[width=3in]{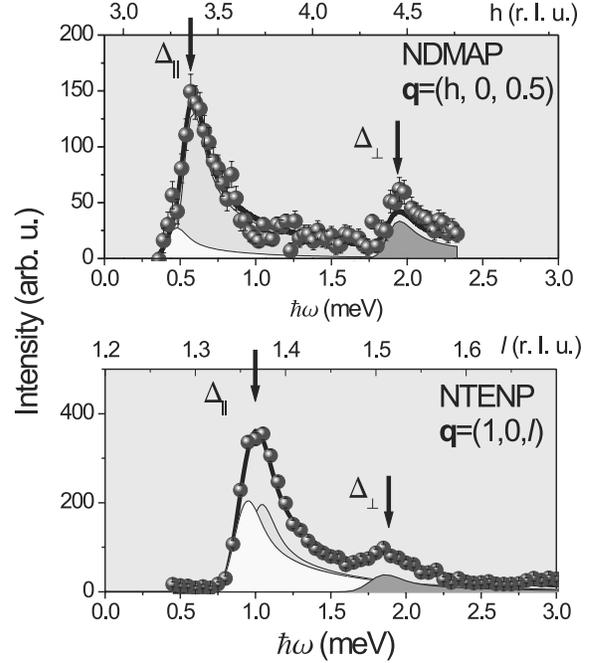}
 \caption{\label{zerofield} Typical neutron spectra collected for the Haldane-gap material NDMAP (top)
 and the $S=1$ dimer-gap system NTENP (bottom) at the respective 1D antiferromagnetic
 zone-centers. NDMAP and NTENP data are from refs.~\cite{Zheludev2001} and \cite{Zheludev2004}, respecively.}
\end{figure}

\paragraph{Structures}The crystal structures of the Haldane-gap material NDMAP and
the bond-alternating chain system NTENP are shown in
Fig.~\ref{struc}. The key features of both are covalent chains of
octahedrally coordinated Ni$^{2+}$ ions. Only nearest-neighbor
in-chain AF interactions are relevant. The coordination geometry
results in a substantial single-ion magnetic anisotropy, with a
magnetic easy plane that is almost perpendicular to the chain
direction. The chains are held together by Van Der Vaals forces,
so inter-chain interactions are negligible. All Ni$^{2+}$ sites
are equaivalent and there is no staggered component to the
Ni$^{2+}$ $g$-factor in either structure. However, in NDMAP all
Ni-Ni exchange pathways along the chains are equivalent, whereas
in NTENP Ni-Ni distances alternate between $d_1=4.28$~\AA\ and
$d_2=4.86$~\AA. The structure of NDMAZ is similar to that of
NDMAP.
\paragraph{Excitations in NDMAP}Both NDMAP and NTENP have a gap in the magnetic excitation
spectrum that is is most prominent in inelastic neutron scattering
spectra. Typical scans collected at the respective 1D AF
zone-centers in both compounds are shown in Fig.~\ref{zerofield}
\cite{Zheludev2001,Zheludev2004-2}. Let us first discuss the
spectrum of NDMAP. There are, in fact, three distinct gap
excitations, two of which are merged into a single peak at lower
energies in Fig.~\ref{zerofield}a. The three peaks correspond to
the three members of the triplet of Haldane gap excitations, split
by magnetic anisotropy. The upper mode is polarized perpendicular
to the magnetic easy-plane, {\it i.e.}, almost parallel to the
spin chains. The asymmetric peak shapes are due to experimental
resolution that is well-characterized. The gap energies for the
three modes are $\Delta_1=0.42(3)$~meV, $\Delta_2=0.52(6)$~meV,
and $\Delta_3=1.9(1)$~meV \cite{Zheludev2001}. Using known
numerical results for anisotropic Haldane spin
chains\cite{Golinelli93,Meshkov93} one can obtain the exchange
constant $J=2.3$~meV ($\overline{\Delta}/J=0.41$) and the
anisotropy parameter $D/J\approx 0.25$, which are in good
agreement with estimates based on high-temperature susceptibility
measurements \cite{Honda98}. The intensities of the three peaks,
apart from the polarization dependence of the inelastic cross
section for unpolarized neutrons, scale almost exactly as
$1/\omega$.
\paragraph{Ground state of NTENP} For NTENP $J=3.4$~meV \cite{Narumi2001},
based on high-temperature measurements. Using the measured gap
energies $\Delta_1=0.42(3)$~meV, $\Delta_2=0.52(6)$~meV, and
$\Delta_3=1.9(1)$~meV one gets $\overline{\Delta}/J=0.4$, i.e.,
which is almost exactly  the same ratio as for a uniform spin
chain \cite{Zheludev2004-2,Regnault2004}. Does NTENP then have a
Haldane ground state with very little alternation of exchange
interactions, or does the gap instead originate from dimerization,
with $\overline{\Delta}/J \approx 0.4$ being just a coincidence?
Deciding which is the case is not trivial, since the only
qualitative distinction between the two possible ground states is
the elusive and unmeasurable string order parameter. Yet, one can
look for quantitative differences. For example, following the
dispersion of magnons in NTENP gives a measure of the
zone-boundary energy $\hbar\omega_\mathrm{ZB}/J=2.2$
\cite{Zheludev2004-2,Hagiwara2005}. This value is smaller than
that for a Haldane spin chain where
$\hbar\omega_\mathrm{ZB}/J=2.7$ \cite{Sorensen94}. In fact, recent
numerical work \cite{Suzuki2003} directly relates
$\hbar\omega_\mathrm{ZB}/J$ to $\delta$ and allows to estimate
$\delta\approx 0.3$ for NTENP. Thus, NTENP is indeed strongly
dimerized and is {\it not} in a Haldane ground state.

Inelastic neutron scattering actually provides a tool for directly
measuring the level of dimerization in a bond-alternating chain.
Applying the Hohenberg-Brickman sum rule \cite{Hohenberg74} for
the first moment of the dynamic structure factor of a
bond-alternating spin chain one gets:
\begin{eqnarray}
\int_{-\infty}^{\infty}(\hbar\omega)S^{\alpha\alpha}(\mathbf{q},\omega)d(\hbar
\omega) =\nonumber\\
 -\frac{4}{3}E_1\sin^2(\mathbf{q}\mathbf{d}_1/2)
 -\frac{4}{3}E_2\sin^2(\mathbf{q}\mathbf{d}_2/2).
 \label{SR2}
\end{eqnarray}
Here $\mathbf{d}_1$ and $\mathbf{d}_2$ are real-space vectors
chosen along the short and long bonds in the chains. The
quantities $E_1=J_1\langle
\mathbf{S}_{2j}\mathbf{S}_{2j+1}\rangle$ and $E_2=J_2\langle
\mathbf{S}_{2j}\mathbf{S}_{2j-1}\rangle$ are ground state {\it
exchange energies} associated with the strong and weak bonds,
respectively. Thus, by measuring the energy-integrated intensities
of magnetic excitations in different Brillouin zones, it was
possible to directly extract the modulation of exchange energy
$\delta'=\frac{E_1-E_2}{E_1+E2}\approx 0.42$ for NTENP
\cite{Zheludev2004-2}. Numerical calculations were then used to
obtain $\delta$ from $\delta'$, reaching the same conclusion:
NTENP is strongly dimerized. Based on a combination of techniques
it was concluded that for NTENP $J_1=2.2$~meV, $J_2=4.7$~meV and
$D\approx 0.3$~meV.

\paragraph{Excitations in NTENP} Let us now return to the spectrum
measured in NTENP at the 1D AF zone center and shown in
Fig.~\ref{zerofield}b. After correcting for the usual polarization
factors, it was found that the intensity of the higher-energy gap
mode ($\Delta=1.9$~meV) in NTENP is anomalously small: at least 4
times weaker than expected from the usual $1/\omega$ scaling
typical of antiferromagnets and obeyed very well in NDMAP. The
physical significance of this observation was recently revealed in
numerical calculations of the excitation spectrum performed using
realistic exchange and anisotropy parameters
\cite{Hagiwara2005,Suzuki2005}. It was shown that in NTENP, unlike
in NDMAP, for the higher-energy-mode there is an open channel of
decay into a pair of lower-energy gap modes. This decay process
leads to a downward renormaliztion of the upper mode's intensity
and gives it a finite lifetime (energy width). An equivalent
interpretation is that the upper mode occurs in the $E-\mathbf{q}$
range occupied by a continuum of 2-magnon excitations. In NTENP
the observed upper mode is very close to the lower continuum
bound.

\begin{figure}
 \includegraphics[width=3in]{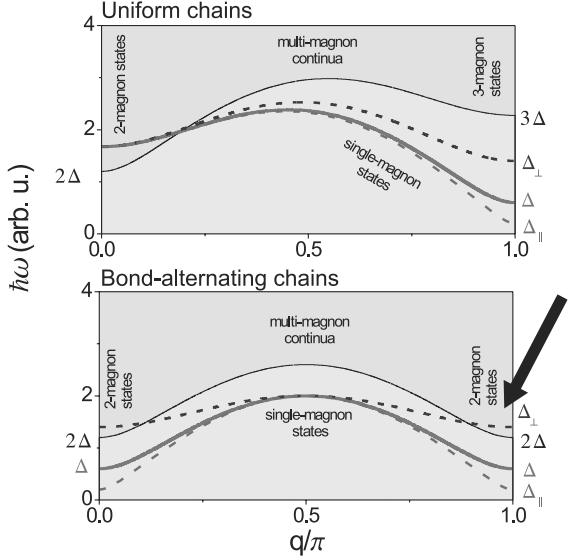}
 \caption{\label{cartoon} A simplified view of the excitation spectra in uniform (top) and
 bond-alternating (bottom) quantum $S=1$ chains. Lines are as explained in the text.}
\end{figure}


The key to understanding this difference in behavior is the
breaking of translational symmetry in bond-alternating spin
chains. Figure~\ref{cartoon} schematically shows the spectra for
the uniform and bond-alternating cases. In both systems the
lowest-energy excitation near $q=\pi$ are a triplet of long-lived
gap excitations at an energy $\Delta$. These are shown in heavy
solid lines, with the assumption that no magnetic anisotropy is
present. In a Haldane spin chain the single-magnon energy becomes
rater large near $q=0$, the spectrum resembling that of a spin
ladder \cite{Barnes94}. As a result, the lowest-energy excitations
near $q=0$ are actually a continuum of 2-magneon states with a gap
of roughly $2\Delta$ \cite{Affleck92}. Not so in a
bond-alternating spin chain where $q=0$ and $q=\pi$ are {\it
equivalent} due to symmetry breaking. Here the dispersion
resembles that of coupled $S=1/2$ dimers and the magnon triplet at
$q=0$ again has a gap of $\Delta$. Returning to $q=\pi$, we see
that in a Haldane chain the lowest-energy excitations above the
magnons are 3-magnon states, starting at a rather high energy of
$3\Delta$ \cite{Essler2000}. In a bond-alternating chain at
$q=\pi$ 2-magnon states are allowed, so the continuum starts at a
lower energy of $2\Delta$. If the magnon branches become split by
anisotropy, as is the case for NDMAP and NTENP, the upper
(out-of-plane) mode at $q=\pi$ moves upwards towards the lower
continuum bound. This is illustrated by dashed lines in
Fig.~\ref{cartoon}. The lower bounds of the continua actually
moves down in the anisotropic case (not shown). In NTENP the upper
mode thus becomes submerged in the low-lying 2-magnon continuum
and can decay into a pair of in-plane magnons. In NDMAP the
continuum gap is larger to begin with, so the upper mode remains
safely below it. Due to limitation on experimental resolution, the
continua can not be clearly identifies, but the effect on the
intensity of the upper mode is easily detected.


\section{High fields}
\begin{figure}
 \includegraphics[width=3in]{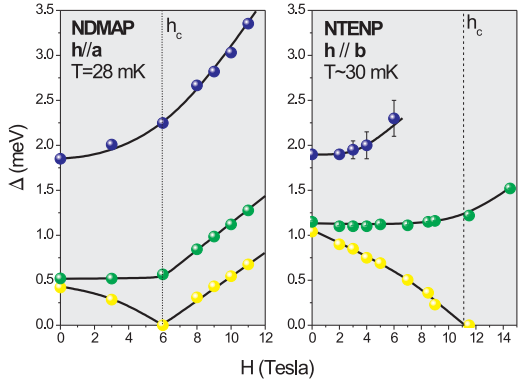}
 \caption{\label{vst} Symbols: measured field dependencies of the gap energies in NDMAP (left,
 Ref.~\protect\cite{Zheludev2003}) and NTENP (right, Ref.~\protect\cite{Hagiwara2005}). Lines are guides for the eye.}
\end{figure}

\paragraph{Zeeman splitting}
Since the gap excitations are an $S=1$ triplet, an external
magnetic field modifies their energies by virtue of Zeeman effect.
Figure~\ref{vst} shows the measured field dependencies of the gaps
in NDMAP and NTENP. Below $H_c$ this behavior remarkably well
accounted for by simple perturbation theory
\cite{Golinelli92,Regnault93}, that considers a 2-nd order mixing
induced between the excited states by the Zeeman term. In the
presence of anisotropy the ground state is also affected, but only
in the next order \cite{Regnault93}, so the effect on the gap
energy is minimal. $H_c$ obtained from perturbation theory is very
close to the actual observed value
\cite{Chen2001,Zheludev2001-2,Zheludev2005,Hagiwara2005}. More
sophisticated theoretical models can achieve an even better
agreement with experiment
\cite{Zheludev2003,Zheludev2004,Hagiwara2003}.

\begin{figure}
 \includegraphics[width=3in]{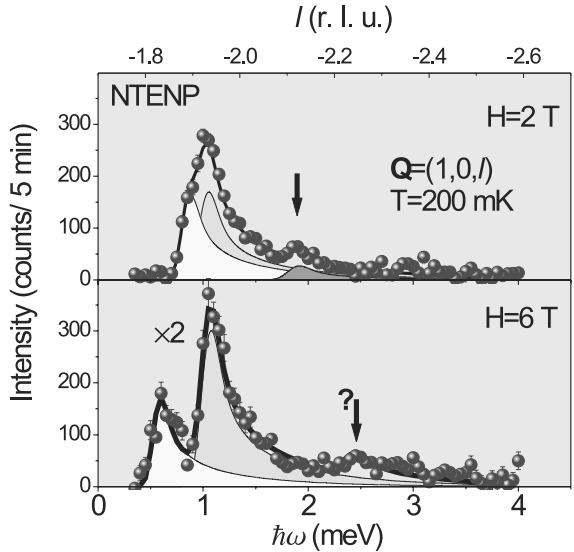}
 \caption{\label{vanishing} Evolution of the excitation spectrum in NTENP at the 1D AF zone-center in
 moderate applied fields. Zeeman splitting of the lower excitation doublet is apparent. The
 upper mode progressively weakens and boadens. It is no longer observable above 6~T. The data
 are from Ref.~\protect\cite{Hagiwara2005}.}
\end{figure}

\begin{figure}
 \includegraphics[width=3in]{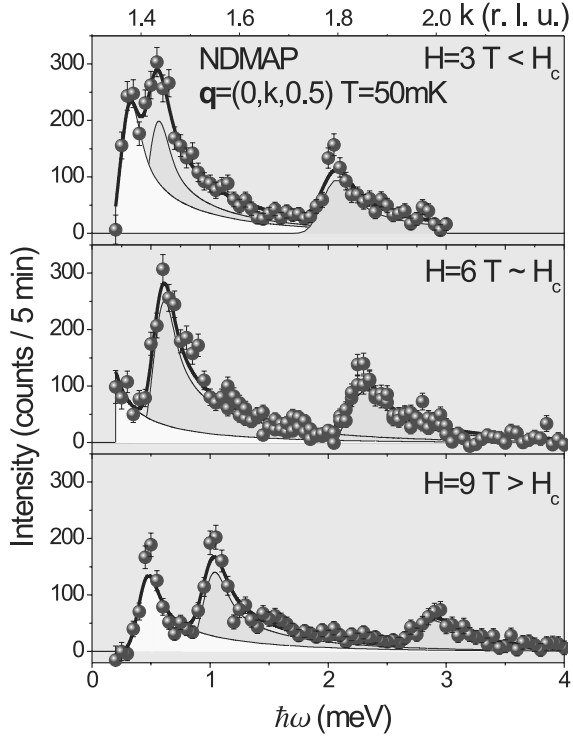}
 \caption{\label{highf1} Evolution of the excitation spectrum of the Haldane-gap system  NDMAP at the 1D AF zone-center upon crossing
 the critical filed $H_c$. All three branches survive in the high-field ordered state. Conventional
 spin wave theory can account for only two transverse-polarized modes.  The data
 are from Ref.~\protect\cite{Zheludev2003}.}
\end{figure}

\begin{figure}
 \includegraphics[width=3in]{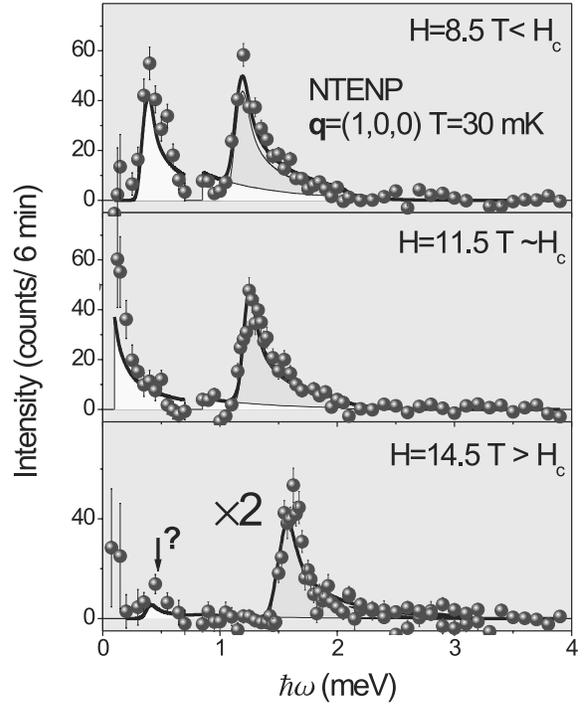}
 \caption{\label{highf2} Evolution of the excitation spectrum of the $S=1$ dimerized system NTENP
 at the 1D AF zone-center upon crossing
 the critical filed $H_c$. Unlike in NDMAP, only one excitation branch survives a an intense sharp excitation
 in the high-field ordered state. The data
 are from Ref.~\protect\cite{Hagiwara2005}.}
\end{figure}

For NDMAP the upper Haldane gap mode persists as a well-defined
sharp excitation all the way to $H_c$ and beyond (see
Fig.~\ref{highf1}). In contrast, in NTENP, where the upper mode is
anomalously weak already at $H=0$, it further weakens and then
totally vanishes well below the transition. This is illustrated in
Fig.~\ref{vanishing} and Fig.~\ref{highf2} that show data from
\cite{Hagiwara2005}. Such behavior is easy to understand within
the hand-waving picture presented in the previous section. With
increasing $H$ the upper
 mode moves further into the 2-magnon
continuum, while the lower bound of the continuum actually moves
down. A larger portion of the 2-particle phase space becomes
available for final decay state of magnons from the upper branch,
which further affects their width and intensity.

\paragraph{Field-induced spin freezing}
Once the gap in the lower mode is driven to zero, the gapped
quantum spin liquid ``freezes''. All high-field experiments on
NDMAP, NDMAZ and NTENP realize the axially asymmentric geometry in
which above the critical field even an isolated chain has
antiferromagnetic long-range order at $T=0$. Weak inter-chain
interactions stabilize this long-range order at non-zero
temperatures \cite{Sakai2000}. A signature of this field-induced
antiferromagnetism is the appearance of new Bragg reflections at
$H>H_{\mathrm{c}}$. Such reflections were observed by means of
neutron diffraction in NDMAP \cite{Chen2001,Zheludev2005}, NDMAZ
\cite{Zheludev2001-2} and NTENP \cite{Hagiwara2005}, and the
approximate structures of the high-filed phases were determined.
In all cases the ordered staggered moment lies perpendicular to
the direction of applied field and within the magnetic easy-plane.
This spin arrangement allows the system to become magnetized: the
ordered moments canttilt slightly in the direction of the field,
to produces a net uniform magnetization. Interestingly, a
relatively small field in excess of $H_c$ can produce a large
staggered moment. For example, in NDMAP, where $H_c=4.5$~T, in
magnetic fields of 7~T the sublattice magnetization is already as
high as 1.2~$\mu_\mathrm{B}$ per site.

\begin{figure}
 \includegraphics[width=3in]{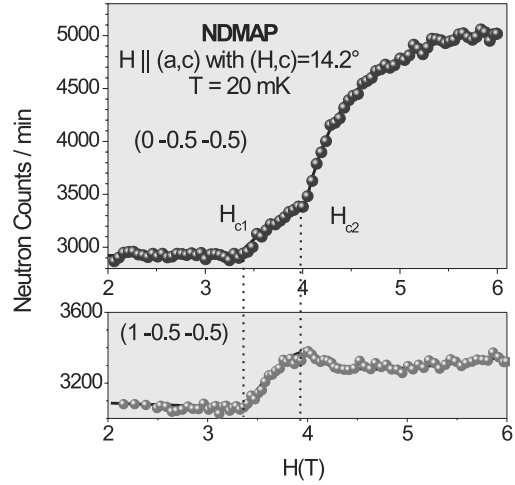}
 \caption{\label{double} Field dependence of fwo magnetic Bragg reflections in NDMAP
 for a field applied at an angle 14$^\circ$ to the major crystal axes. Two consecutive
 spin-freezing transitions are apparent. The data
 are from Ref.~\protect\cite{Zheludev2005}.}
\end{figure}

As far as static properties are concerned, there are no
qualitative differences between magnetized Haldane versus
dimerized spin chains. Nevertheless, two peculiarities  of spin
freezing in NDMAP are worth mentioning. First of all, in NDMAP
there are two sets of AF $S=1$ chains that are totally decoupled
due to geometric frustration of exchange interactions. The local
anisotropy axes of all magnetic ions tilted by $\alpha=16^\circ$
relative to the crystal $c$ axis. The tilt directions are {\it
opposite} for spins belonging to the two different sets of chains.
As a result, when the field is applied in a direction other than
along the principal crystallographic axes, the angles between
local anisotropy axes and field are {\it different} for the two
sets of chains. Spin freezing then occurs first only in one set of
chains at a critical field $H_\mathrm{c1}$. The other chains
remain gapped and disordered up to a second critical field
$H_\mathrm{c2}>H_\mathrm{c1}$, at which they will also order. This
two-stage spin freezing results in a peculiar field dependence of
magnetic Bragg intensities as shown in Fig.~\ref{double}, and
explained in detail in Ref.~\cite{Zheludev2005}.

Another interesting feature of spin freezing NDMAP is due to that
non-frustrated inter-chain interactions span over at least 18\AA~
along the $a$ axis. They are extremely weak and presumably of
dipolar origin. As these interactions are crucial to establishing
true long-range order, the degree of ordering becomes dependent on
the filed direction. In most geometries for NDMAP a 3D-ordered
state is observed at high fields, just as for NDMAZ and NTENP.
However, in experiments where the field was applied perpendicular
to the chains, one instead observes {\it two}-dimensional spin
freezing \cite{Chen2001}. Sufficiently long-range AF spin
correlations are established in layers parallel to the $(b,c)$
plane, but such layers remain uncorrelated along the $a$
direction. This unusual {\it anisotropic correlated spin glass}
state is characterized by Bragg {\it rods} (as opposed to Bragg
{\it peaks}) parallel to the $a^\ast$ axis.

\paragraph{Dynamics of quantum spin solids}
Even though at high fields the ground states of the Haldane gap
material NDMAP and the dimerized system NTENP are characterized by
long-range antiferromagnetic order as in ``conventional'' magnets,
their excitation spectra are {\it nothing} like that predicted by
semiclassical spin wave theory (SWT). Figures~\ref{highf1} and
~\ref{highf2} are a comparison of energy scans measured at the 1D
AF zone-centers in NDMAP in NTENP at 3~T below the critical field,
almost exactly at the quantum phase transition point, and at 3~T
above.

The most striking feature of the spectrum of NDMAP is the
persistence of {\it three} sharp gap above $H_\mathrm{c}$
\cite{Zheludev2003,Zheludev2004}. The gap in the lower mode closes
at the transition point but promptly re-opens in higher fileds.
SWT can account for only {\it two} excitation branches. Indeed,
conventional spin waves represent a precession of the ordered
moments around their equilibrium orientation ($z$ axis), and are
thus necessarily polarized in the {\it transverse} direction. Two
independent precessions, with $S_z=\pm1$ are all there is. In
NDMAP the polarizations of magnetic excitations above the critical
field have not been identified to date. Nevertheless, it is clear
that one of the three branches must be a {\it longitudinal} mode
with $S_z=0$, representing fluctuation of the {\it magnitude} of
the ordered moment. Where SWT is fully inadequate, a number of 1D
field-theoretical models work rather well. These are typically
based on mappings of the original Heisenberg Hamiltonian with
anisotropy onto the quantum non-linear $\sigma$-model, its
descendent the so-called $\phi^4$-model
\cite{Affleck90,Affleck91,Mitra94}, Majorana fermions
\cite{Tsvelik90,Essler2004}, or certain dimer models
\cite{Zheludev2003,Hagiwara2003,Zheludev2004}. It is important to
emphasize that the re-opening of the gap in all cases is a {\it
direct consequence} of the breaking of axial symmetry by a
combination of easy-plane and in-plane anisotropies, or by the
field configuration in particular experiments. In the axially
symmetric case one expects a gapless high-filed phase.

The high-field excitation spectrum in NTENP is qualitatively
different from that in NDMAP. As can bee seen in
Fig.~\ref{highf2}, in the dimerized system there is only {\it one}
distinct long-lived excitation above $H_\mathrm{c}$, a descendant
of middle member of the original triplet at $H=0$. At low
energies, where in NDMAP one sees a new gap excitation, in NTENP
there is very little scattering at $H>H_c$. Scans in different
Brillouin zones confirm that the effect is not due to experimental
polarization selectivity. A clue to what is going on is again
found in numerical simulations for NTENP
\cite{Hagiwara2005,Suzuki2005}. It appears that above $H_c$ not
just the upper mode, but also the lowest-energy excitations are
actually diffuse multi-magnon continua, rather than sharp
single-magnon states. A very recent study\cite{Regnault2005}
confirmed this picture and detected both continua.

Are the striking differences in the high-filed spin dynamics of
NTENP and NDMAP related to the different nature of their ground
states? In answering this question the future development of
analytical models for dimerized integral spin chains will be very
helpful.

\section{New directions}
\paragraph{Breakdown of the single-mode approximation}
Experiments on NTENP have demonstrated that even in the
long-wavelength limit (near the 1D AF zone-center), long-lived
single-particle excitations are only part of the story.
Excitations with short lifetimes and entire diffuse excitation
continua may constitute a substantial or even dominant fraction of
the total spectral weight. Recent theoretical work
\cite{Essler2004} predicts that in Haldane spin chains one  can
expect similar phenomena. In particular, at some crossover field
$H'<H_c$ the gap modes should become short-lived. The energy
resolution of previous experiments on NDMAP was not sufficient to
observe this effect. However, recent high-filed neutron scattering
studies on NENP contain clear evidence of broadening in at least
two of the Haldane gap branches around $H_c$
\cite{Zaliznyak-unpublished}. Re-investigating the excitation
lifetime issue in NDMAP is clearly worthwhile.

\paragraph{Isotropic $S=1$ chains}
As mentioned in the introduction, the physics of the high-field
phase transition in the isotropic model is expected to be totally
different from that found in the strongly anisotropic NDMAP and
NTENP systems. Until recently, no good model $S=1$ chain compounds
with gap energies suitable for high field inelastic neutron
scattering experiments and weak enough inter-chain interactions
could be identified. A new material that seems to be almost ideal
in this respect is (CH$_3$)$_2$CHNH$_3$CuCl$_3$ (abbreviated as
IPA-CuCl$_3$). This compound is composed of $S=1/2$ ladders that
have {\it ferromagnetic} rungs, which makes them  equivalent to
Haldane $S=1$ chains \cite{Masuda2005-LT}. Anisotropy affects are
negligible due to the nature of the magnetic Cu$^{2+}$ ions. Bulk
measurements revealed a field-induced Bose condensation of magnons
and a quantum phase transition with $H_c\simeq 9$~T, with a
remarkably isotropic $H-T$ phase diagram \cite{Manaka98}. High
field neutron scattering experiments on deuterated IPA-CuCl$_3$
are planned for the near future.

\paragraph{Incommensuration in anisotropic chains}
In the presence of axial symmetry the spin correlations above
$H_c$ are expected to be incommensurate. As discussed above, in
NTENP and NDMAP the magnetized phases are commensurate due to
strong anisotropy effects. Only a hint of incommensurate spin
correlations was detected in NDMAP, at temperatures high enough to
destroy long range order \cite{Zheludev2002}. Recent theories
\cite{Wang2003,Essler2004} predict that even anisotropic models
can acquire incommensurate order above some critical
incommensuration field $H^\ast>H_c$. For NDMAP, with a field
applied along the chain axis, a crude estimate yields
$H^\ast\approx 15$~T. High-field calorimetric experiments recently
discovered some sort of phase transition in NDMAP at $H\approx
14$~T \cite{Tsujii2005}. It was speculated that it may be related
to incommensuration. Upcoming neutron diffraction experiments will
test this conjecture.

\section{Conclusion}
The most interesting physics is usually  to be found in the {\it
complex} behavior of {\it simple} model systems. Integral spin
chains in external magnetic fields are a beautiful example of
this.

\begin{theacknowledgments}
I would like to thank all those who graciously allowed me to
collaborate with them on the NDMAP, NDMAZ and NTENP projects: T.
Barnes (ORNL), C. Broholm (Johns-Hopkins University), Y. Chen
(NIST), J. H. Chung (NIST), R. Feyerherm (HMI), B. Grenier (CEA
Greoble), M. Hagiwara (Osaka University), Z. Honda (Saitame
University), K. Kakurai (JAERI), K. Katsumata (RIKEN Harima
Institute), Y. Koike (JAERI), A. K. Kolezhuk (Institute of
Magnetism, Kiev, Ukraine), D. Mandrus (ORNL), T. Masuda (ORNL), N.
Metoki (JAERI), H.-J. Mikeska (Universit\"{a}t Hannover),  T.
Papenbrock (ORNL), S. Park (NIST), K. Prokes (HMI), Y. Qiu (NIST),
L. P. Regnault (CEA Grenoble), E. Ressouche (CEA Greoble), B.
Sales (ORNL), S. M. Shapiro(BNL), A. Stunault (ILL), S. Suga
(Osaka University), T. Suzuki (Osaka University), P. Vorderwisch
(HMI).
Work at ORNL was carried out under DOE
Contract No. DE-AC05-00OR22725.
\end{theacknowledgments}


\end{document}